\documentclass{elsart}
\usepackage{graphicx}
\usepackage{epstopdf}
\usepackage{amssymb}
\begin{document}
%
%
\title{Correlated  $\Lambda d$ pairs from the
$K^{-}_{stop} A \rightarrow \Lambda d \,A'$ reaction} 
\begin{frontmatter}
\centering{\bf FINUDA Collaboration}
\author[polito]{M.~Agnello}, 
\author[victoria]{G.~Beer},
\author[lnf]{L.~Benussi},
\author[lnf]{M.~Bertani},
\author[korea]{H.C. Bhang},
\author[lnf]{S.~Bianco},
\author[unibs]{G.~Bonomi},
\author[unitos]{E.~Botta}, 
\author[units]{M. Bregant},
\author[unitos]{T.~Bressani},
\author[unitos]{S.~Bufalino},
\author[unitog]{L.~Busso},
\author[infnto]{D.~Calvo},
\author[units]{P.~Camerini},
\author[enea]{M. Caponero},
\author[infnto]{P.~Cerello},
\author[uniba]{B.~Dalena},
\author[unitos]{F.~De~Mori},
\author[uniba]{G.~D'Erasmo},
\author[uniba]{D.~Di~Santo}, 
\author[unibo]{R. Don\`a},
\author[uniba]{D. Elia},
\author[lnf]{F.~L.~Fabbri},
\author[unitog]{D.~Faso}, 
\author[infnto]{A.~Feliciello}, 
\author[infnto]{A.~Filippi}, 
\author[infnpv]{V.~Filippini\thanksref{deceased}},
\author[uniba]{R. Fini}, 
\author[uniba]{M.~E.~Fiore}, 
\author[tokyo]{H.~Fujioka},
\author[lnf]{P.~Gianotti},  
\author[infnts]{N.~Grion}, 
\author[lnf]{O.~Hartmann}, 
\author[jinr]{A.~Krasnoperov}, 
\author[lnf]{V.~Lucherini},
\author[uniba]{V. Lenti},
\author[infnba]{V. Manzari},  
\author[unitos]{S.~Marcello}, 
\author[tokyo]{T.~Maruta}, 
\author[teheran]{N.~Mirfakhrai}, 
\author[inaf]{O.~Morra}, 
\author[kek]{T.~Nagae}, 
\author[triumf]{A.~Olin},
\author[riken]{H.~Outa}, 
\author[lnf]{E.~Pace}, 
\author[lnf]{M. Pallotta},
\author[uniba]{M.~Palomba}, 
\author[infnba]{A.~Pantaleo}, 
\author[infnpv]{A.~Panzarasa}, 
\author[infnba]{V.~Paticchio}, 
\author[infnts]{S.~Piano\thanksref{corresponding}}, 
\author[lnf]{F.~Pompili},  
\author[units]{R.~Rui}, 
\author[uniba]{G.~Simonetti}, 
\author[korea]{H.~So}, 
\author[jinr]{V. Tereshchenko},
\author[lnf]{S.~Tomassini}, 
\author[kek]{A. Toyoda},
 \author[infnto]{R.~Wheadon}, 
\author[unibs]{A.~Zenoni}
\thanks[corresponding]{corresponding author. E-mail: stefano.piano@ts.infn.it; 
Fax:++39.040.5583350.}
\thanks[deceased]{deceased}
\address[polito]{Dip. di Fisica Politecnico di Torino, Corso Duca degli 
Abruzzi Torino, Italy, and INFN Sez. di Torino, via P. Giuria 1 Torino, Italy}
\address[victoria]{University of Victoria, Finnerty Rd.,Victoria, Canada}
\address[lnf]{Laboratori Nazionali di Frascati dell'INFN, via E. Fermi 40 
Frascati, Italy}
\address[korea]{Dep. of Physics, 
Seoul National Univ., 151-742 Seoul, South Korea}
\address[unitos]{Dipartimento di Fisica Sperimentale, Universit\`a di
Torino, via P. Giuria 1 Torino, Italy, and INFN Sez. di Torino, 
via P. Giuria 1 Torino, Italy} 
\address[units]{Dip. di Fisica Univ. di Trieste, via Valerio 2 Trieste, 
Italy and INFN, Sez. di Trieste, via Valerio 2 Trieste, Italy}
\address[unitog]{Dipartimento di Fisica Generale, Universit\`a di
Torino, via P. Giuria 1 Torino, Italy, and INFN Sez. di Torino, 
via P. Giuria 1 Torino, Italy} 
\address[infnto]{INFN Sez. di Torino, via P.  Giuria 1 Torino, Italy}
\address[enea]{ENEA, Frascati, Italy}
\address[uniba]{Dip. di Fisica Univ. di Bari, via Amendola 179 Bari, 
Italy and INFN Sez. di Bari, via Amendola 179 Bari, Italy }
\pagebreak
\address[unibo]{Dipartimento di Fisica, 
Universit\`{a} di Bologna, via Irnerio 46, Bologna, Italy and 
INFN, Sezione di Bologna, via Irnerio 46, Bologna, Italy}
\address[infnpv]{INFN Sez. di Pavia, via Bassi 6 Pavia, Italy}
\address[tokyo]{Dep. of Physics Univ. of Tokyo, Bunkyo Tokyo 
113-0033, Japan}
\address[infnts]{INFN, Sez. di Trieste, via Valerio 2 Trieste, Italy}
\address[jinr]{Joint Institute for Nuclear Research (JINR), Dubna, Russia}
\address[teheran]{Dep of Physics Shahid Behesty Univ., 19834 Teheran, Iran}
\address[inaf]{INAF-IFSI Sez. di Torino, C.so Fiume, Torino, Italy
and INFN Sez. di Torino, via P. Giuria 1 Torino, Italy} 
\address[kek]
{High Energy Accelerator Research Organization (KEK), Tsukuba, Ibaraki
305-0801, Japan}
\address[triumf]{TRIUMF, 4004 Wesbrook Mall, Vancouver BC V6T 2A3, Canada} 
\address[riken]{RIKEN, Wako, Saitama 351-0198, Japan}
\address[infnba]{INFN Sez. di Bari, via Amendola 179 Bari, Italy }
\address[unibs]{Dip. di Ingegneria Meccanica e Industriale, 
Universit\`a di Brescia, via Branze 38 Brescia, Italy and 
INFN Sez. di Pavia, via Bassi 6 Pavia, Italy}
\end{frontmatter} 
%
%
\vspace{-0.6cm}
{\small\bf Abstract:}
{\small
Correlated $\Lambda d$ pairs emitted after the absorption of negative 
kaons at rest $K^{-}_{stop}A\rightarrow \Lambda d\,A'$ in light nuclei 
$^6Li$ and $^{12}C$ are studied. $\Lambda$-hyperons and deuterons are 
found to be preferentially emitted in opposite directions. The $\Lambda d$ 
invariant mass spectrum of $^6Li$ shows a bump whose mass is 
3251$\pm$6 MeV/c$^2$.  The bump mass (binding energy), width and yield 
are reported. The appearance of a bump is discussed in the realm of the 
[$\overline{K}3N$] clustering process in nuclei. The experiment was 
performed  with the FINUDA spectrometer at DA$\Phi$NE (LNF).
}

\begin{picture}(0,0)
\linethickness{0.5pt}
\put(-185,12){\line(1,0){478}}
\end{picture}
\normalsize
%
%
\vspace{0.5cm}
\begin{center}
{\bf I. INTRODUCTION}
\end{center}
In this letter we investigate the invariant mass spectra of correlated 
$\Lambda d$ pairs, which are produced by the kaon absorption reaction 
$K^{-}_{stop} A\rightarrow\Lambda d\, A'$. The nuclei ($A$) examined 
are $^6Li$ and $^{12}C$ and the $\Lambda$-hyperon and deuteron  are 
the reconstructed particles in the reaction final state. The present 
study follows an earlier $\Lambda p$ survey on light and medium-light 
nuclei made by the FINUDA collaboration \cite{expt:FINUDA0}, which 
supported the view that negative kaons gather nucleons to form bound 
systems; i.e., $K^-_{stop}pp\rightarrow [K^-pp]\rightarrow\Lambda p$. 
The present discussion about the dynamics of [$\overline{K}3N$] clusters 
in $A$ is pursued by studying the $A(K^{-}_{stop},\Lambda d)A'$  reaction 
channel.
   
A [$\overline{K}3N$] cluster was found in the course of a search for 
neutrons emitted from the $K^-_{stop}$$^4He\rightarrow  nA'$ absorption 
reaction, where  $A'$ is the residual nucleus \cite{expt:iwasaki}. These
neutron data, which result from semi-inclusive measurements, show a bump 
in the  excitation spectrum in the range 450-500 MeV/c. The resulting 
missing mass spectrum also has this bump interpreted by the authors as a 
[$\overline{K}3N$] bound state, named the {\em strange tribaryon $S^+$} 
\cite{expt:iwasaki}. The measured $S^+$ mass and width are 3140.5 MeV/c$^2$ 
and $<$21.6 MeV/c$^2$, respectively. The theoretical explanation of the 
$S^+(3140)$ as the distinctive pattern of a deeply-bound $\overline{K}$ 
state \cite{theor:akaishi} was contentious; in fact, theory requires an 
uncommonly deep potential with a small imaginary part to reproduce the 
data \cite{theor:oset1}. The [$\overline{K}3N$] core density, which may 
be as high as 5 times nuclear density, was also contentious. 
 
Experimental investigation of the existence of $K^-$-bound nuclear 
states began several years ago \cite{expt:FINUDA0,expt:iwasaki}. 
Theoretical studies were started earlier. The reality of these states 
was never disputed, although the values of their binding energy and 
width strongly depend on both the model and the parameters used. For 
example, the $\overline{K}$-nuclear potential is predicted to range 
from deep (150-200 MeV \cite{theor:friedman}) to shallow (50-60 MeV
\cite{theor:baca}). Widths, also predicted to vary, are critical for 
the detection of $K^-$-nuclear states.   

Proof of the existence of strange heavy baryons or the actuality of 
nuclear $\overline{K}$ bound states, whether discrete  or not, cannot 
simply rely on (semi-)inclusive measurements. In fact, the erroneous 
interpretation of a bump observed in the energy distribution of protons 
from the $K^-_{stop}$$^4He\rightarrow p\,A'$ reaction led to prediction
of the neutral strange tribaryon $A'\equiv S^0(3115)$\cite{expt:iwasaki1}.  
The data discussed in this article exploit the capability of the FINUDA 
spectrometer to fully reconstruct the $A(K^{-}_{stop}, \Lambda d)A'$
reaction with $A'$ undetected, thereby yielding a quasi-complete kinematic 
picture of the detected events.
\begin{center}
{\bf II. THE EXPERIMENTAL METHOD}
\end{center}
The search for [$\overline{K}3N$] clusters makes use of the absorption
reaction  $K^-_{stop} \,A\rightarrow\Lambda(1116) d\,A'$, where $A$ is
either  $^{6}Li$ or $^{12}C$ and  $A'$ is  a system of (A-3) nucleons 
not necessarily bound. The experimental method is briefly explained in 
this letter; however, further details can be found in Refs.
\cite{expt:FINUDA2,spectr:FINUDA0,spectr:FINUDA}.   

At the DA$\Phi$NE collider at LNF, 510 MeV electron/positron collisions 
produce 16.1$\pm$1.5 MeV kaons as a by-product of the multistep 
process $e^+e^-\rightarrow\Phi(1020)\rightarrow K^+K^-$ (B.R.$\sim$50\%). 
The kaons slow down as they traverse some of the sensitive layers of FINUDA 
until they come to rest within solid targets as thin as 0.213 g/cm$^2$ 
for $^6Li$ and 0.295 g/cm$^2$ for $^{12}C$. Both kaons and their decay 
or reaction products, $\mu$'s, $\pi$'s, $p$'s and $d$'s, are analyzed in 
FINUDA. The magnetic spectrometer has cylindrical geometry, centered on 
the $\Phi(1020)$-production volume. The magnetic field was set at 1.0 T. 
The innermost sensitive layer, (TOFINO) \cite{FINUDA:02}, is a segmented 
detector made of plastic scintillator, which is optimized for starting the 
time-of-flight system. TOFINO is followed by two layers of double-sided 
silicon strip detectors, which are used  for both localization and 
identification of charged particles (ISIM and OSIM)\cite{FINUDA:03}. The 
(eight) targets are accommodated between ISIM and OSIM. Next, two layers of 
drift chambers (LMDC) identify and locate charged particles \cite{FINUDA:04}. 
A stereo-arrangement of straw tubes is the last sensitive tracking layer 
located within the magnetic field (ST) \cite{FINUDA:05}. This stereo device 
is the last layer used for particle localization. The outermost  layer of 
FINUDA consists of a segmented detector, TOFONE, made of rectangular slabs 
of plastic scintillator \cite{FINUDA:06}. TOFONE is designed as the stop 
for the FINUDA time-of-flight measurements and also measures the energy 
released by charged particles. 

%
%
\begin{figure}[t,c,b]
 \centering
  \includegraphics*[angle=0,width=0.7\textwidth]
   {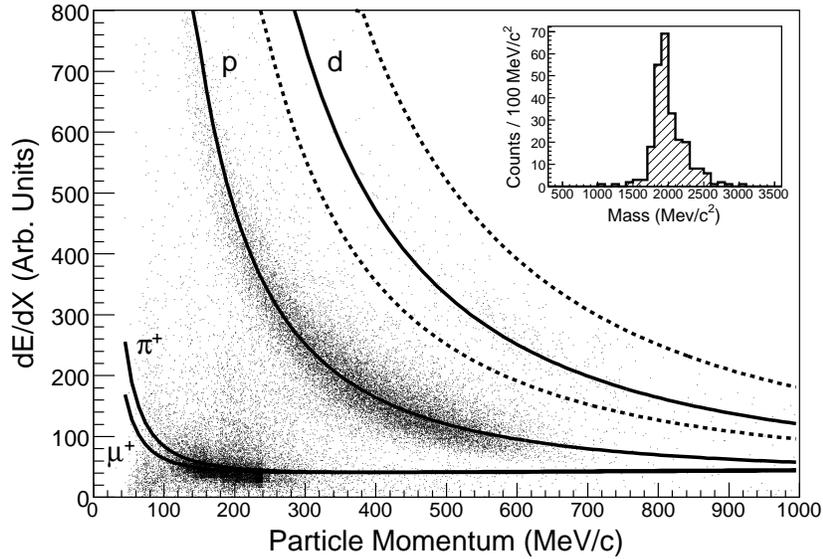}
    \caption{\footnotesize  $dE/dx$ vs momentum diffusion plot of $\mu$'s, 
      $\pi$'s, $p$'s and $d$'s. Full lines, Bethe-Bloch functions. The 
      dashed lines delimit the full-width-at-tenth-maximum region of 
      deuterons. Inset, mass distribution of the particles falling into 
      the deuteron region. More details are given in the text.}
\end{figure}      
The charged particles involved in the absorption (or decay) process are 
mass-identified by  specific energy deposit ($dE/dx$) in some of the  
layers of the spectrometer. In fact, a coherent response from a minimum of 
3 $dE/dx$ layers is required to identify (ID) a particle as a pion, a proton  
or a deuteron. Fig. 1 shows the $dE/dx$-response of OSIM (scatterplots) 
and the Bethe-Bloch functions (full lines) for $\mu$'s, $\pi$'s, $p$'s and 
$d$'s, as a function of the particle momenta.  These particles, the 
result of minimum-bias triggers, are not related to any specific kaon 
absorption reaction channel.  The dashed lines define the deuteron region, 
which is chosen by means of a full-width-at-tenth-maximum (FWTM) criterion. 
To check the validity of this criterion, the mass of the particles falling 
into the deuteron  region is independently determined  by  measuring their 
time-of-flight between TOFINO and TOFONE. The particle mass distribution is 
shown in  the inset of Fig. 1. It appears as a peak centered at a mass value 
of $\approx$1880 MeV with a FWTM strength that accounts for more than 96\% 
of the total strength. 

%
%
\begin{figure}[t,c,b]
 \centering
  \includegraphics*[angle=0,width=0.4\textwidth]
   {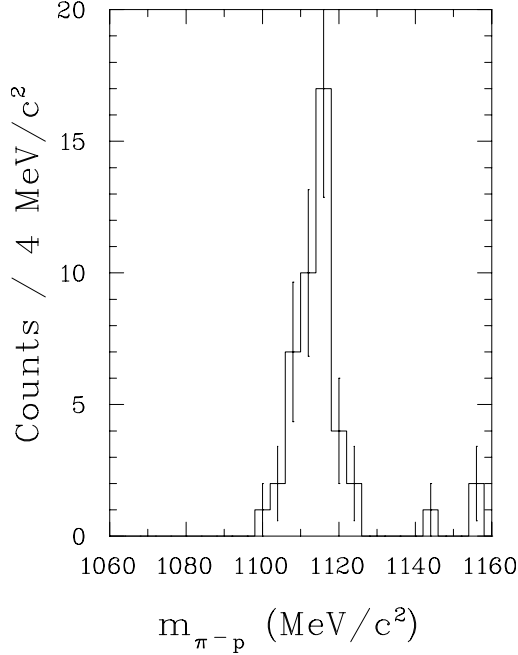}
    \caption{\footnotesize Invariant mass distribution of $\pi^- p$ 
      pairs for $^6Li$. The $\pi^- p$ events forming the histogram 
      are in coincidence with deuterons.}
 \end{figure}      
The initial selection of $\pi^- pd$ events relies on the particle ID 
of FINUDA. Such events are seldom accompanied by other charged particles. 
Once a $\pi^- p d$  event is identified, the $\Lambda(1116) d$ channel is 
entirely reconstructed as well as the track of the stopped negative kaons. 
The topology of the $\Lambda(\rightarrow p\pi^-)d$ events, when combined 
with the selective particle ID of FINUDA, provides distributions nearly 
free from accidental as well as combinatorial background. As an example, 
Fig. 2 shows the $m_{\pi^- p}$ distribution of $^6Li$. The uniform 
distribution of background events over the mass ranges 1080-1102 and 
1128-1150 MeV around the $\Lambda$ peak makes it possible to assess the 
number of such events inside the peak itself. The background events are 
2.2$\pm$0.9 out of a total of 46 $\Lambda$ events, which were produced by 
3.38$\times 10^6 K^-_{stop}$. $\Lambda$ hyperons are detected from 140 
MeV/c (threshold) up to 700 MeV/c, whereas deuterons from the $\Lambda d$ 
channel are analyzed starting from 300 MeV/c (threshold) up to momenta of 
about 800 MeV/c. Both $\Lambda$'s and $d$'s are measured with a resolution  
$\Delta p/p<$2\%. Opening angles in the full range from  
$0^\circ\le\Theta_{\Lambda d}<180^\circ$ were  measured. Spectra  are 
corrected for the spectrometer acceptance,  which also includes the 
event reconstruction efficiency. The acceptance for the $\Lambda d$ 
invariant mass ($m_{\Lambda d}$) increases nearly linearly between 3100 
and about 3340 MeV/c$^2$ (curve not shown). The error bars reflect the 
systematic uncertainty due to the evaluation of the FINUDA acceptance. 
The dominant statistical uncertainty is summed in quadrature with this 
systematic uncertainty in the following analysis,  and overall error bars 
are  reported  in Figs. 3, 4 and 5.
\begin{center}
\bf {III. THE RESULTS}
\end{center}
 Fig. 3 shows the $\Lambda d$ invariant mass  for the 
$^6Li$ target. We observed a bump at $m_{\Lambda d}\sim$3250 MeV/c$^2$ 
and the purpose of the present analysis is to try to understand its 
nature. To this end, the $m_{\Lambda d}$ distribution (thin-line 
histogram) is compared with the phase-space behavior of $\Lambda d$ 
pairs from the $K^-_{stop}$$^6Li \rightarrow \Lambda d\;3N$ absorption 
reaction. The following three reaction channels were investigated: (1) 
$\Lambda d$$nnp$, (2) $\Lambda d$$nd$ and (3) $\Lambda d$$t$. The
$\Lambda d$ pairs from (1) (dashed curve) have an invariant mass 
distribution which shows negligible strength above 3210 MeV/c$^2$; 
therefore, these pairs barely contribute to the bump at 3250 MeV/c$^2$. 
The small-dot curve depicts the $m_{\Lambda d}$ phase space of channel 
(2). The mass distribution of channel (3) (dot-dash curve) also extends
%
%
\begin{figure}[b,c,t]
 \centering
  \includegraphics*[angle=90,width=0.8\textwidth]
   {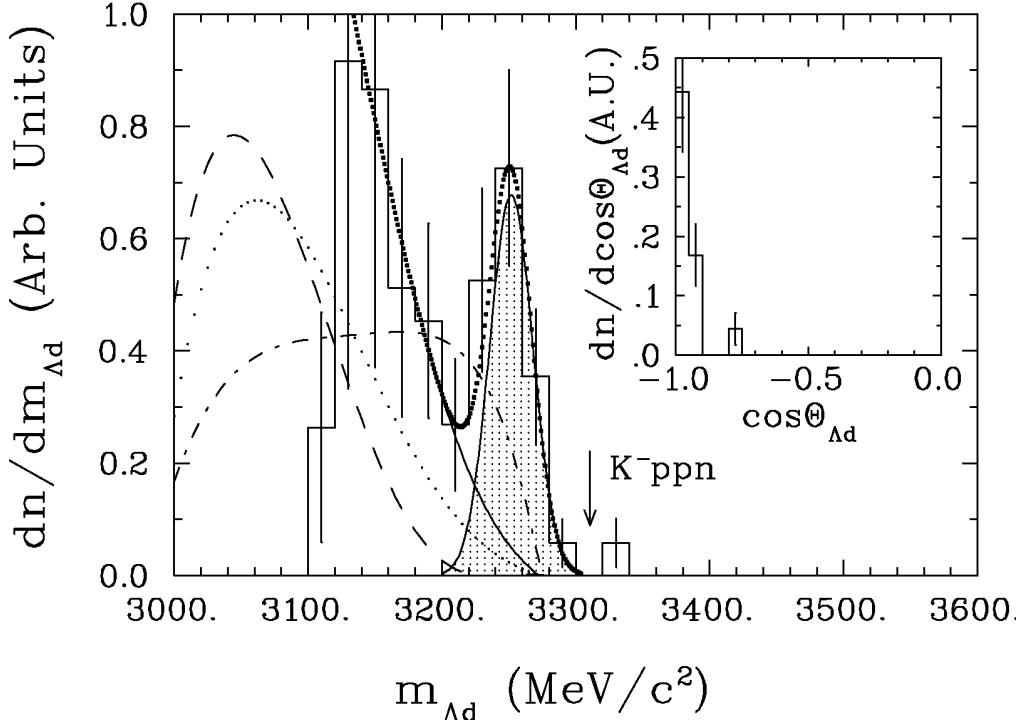}
    \caption{\footnotesize Invariant mass distribution of $\Lambda d$ 
    pairs from the $^6Li(K^-_{stop}, \Lambda d)3N$ reaction (full-line
    histogram). Dash-line, small-dot-line and dash-dot-line curves, phase 
    space simulations of the (1) $\Lambda d$$nnp$, (2) $\Lambda d$$nd$ 
    and (3) $\Lambda d$$t$ reaction channels, respectively. Big-dot-line 
    curve, data fitted with a linear combination of channels (1), (2) and 
    (3) and a Gaussian function (gray-fill curve). Full-line curve, 
    simulated background. The arrow indicates the overall mass of the 
    unbound $K^-ppn$ system. Inset, cos$\Theta_{\Lambda d}$ distribution 
    of the events populating the bump at  3250 MeV/c$^2$.}
\end{figure}      
through the mass range of the data. These three distributions were 
arbitrarily normalized to the data. The phase space simulations indicate 
that the $m_{\Lambda d}$ threshold is slightly below 3000 MeV/c$^2$, a 
mass limit which cannot be probed by FINUDA. This is due to the low 
acceptance of the apparatus below 3100 MeV/c$^2$ combined with poor 
statistics. Consequently, the peak shape of the spectrum at 
$\approx$3140 MeV/c$^2$ may be an artifact. 

$\Sigma^0$ hyperons may also feed the $\Lambda d$ channel via  
$\Sigma^0\rightarrow\Lambda\gamma$ decays. In this case, the energy 
taken away by the $\gamma$ rays ($T_{\gamma}$=74 MeV) shifts  the 
$m_{\Lambda d}$ endpoints of channels (1), (2) and (3) lower, rendering 
their contribution to the $m_{\Lambda d}$ distribution above 3100 MeV/c$^2$
less significant. In addition, the  branching ratios for $K^-_{stop}$ in 
$^4He$ are  2.3\% to $\Sigma^0$ and 9.4\% to $\Lambda$ \cite{expt:katz}, 
which further lessens the role of the $\Sigma^0$  to the formation of 
$m_{\Lambda d}$ in the range of interest. Thus, the excitation and decay 
of intermediate $\Sigma^0$'s  is neglected in the present analysis. 

$\Lambda d$ pairs may also be produced by multi-step processes such 
as (4) where the negative kaon is absorbed by a single nucleon 
$K^-_{stop} N [A-1] \rightarrow \Lambda \pi [A-1]$ followed by a pion 
final state interaction (FSI) $\pi [A-1]\rightarrow d[A-3]$, or (5) 
where a $K^-_{stop}$ is absorbed by a pair of nucleons 
$K^-_{stop} NN [A-2] \rightarrow \Lambda N [A-2]$ and a $N$ pick-up 
reaction $N [A-2] \rightarrow d [A-3]$ follows. Both processes were 
modeled by constraining the interacting $N$ and $NN$ nucleons to obey 
Fermi motion. In the multi-step process (4), the momentum of $\Lambda$'s 
($p_{\Lambda}$) is below 250 MeV/c, while the measured $p_{\Lambda}$ is 
in the range $\approx$450-700 MeV/c. Kinematic constraints also disfavor 
process (5); in fact, the phase space of 
$|\vec{p}_{\Lambda}$ + $\vec{p}_d|$ vs $m_{\Lambda d}$ 
of (5) only slightly overlaps the experimental data.

The smooth behavior of reaction channels (1) $\Lambda d$$nnp$, 
(2) $\Lambda d$$nd$ and (3) $\Lambda d$$t$  is unable to explain 
the structure of the bump at the tail of the measured $m_{\Lambda d}$ 
distribution. To obtain its position ($m_{\Lambda d}$), width 
($\Gamma_{\Lambda d}$) and yield ($Y_{\Lambda d}$), the $\Lambda d$ 
invariant mass distribution was fitted to a linear combination of the 
following distributions: the phase space of channels (1), (2) and (3), 
and a Gaussian function ($G$) whose parameters are left free. The result 
is the big-dot curve which overlaps the full-line curve when the Gaussian 
strength is set to zero. In this approach, the  events underneath the 
full-line curve are the background events. Finally, the Gaussian 
distribution is represented by the grey-fill curve whose parameters are 
$m_{\Lambda d}$=3251$\pm$6 MeV/c$^2$, $\Gamma_{\Lambda d}$=36.6$\pm$14.1 
MeV/c$^2$ and  $Y_{\Lambda d}$=(4.4$\pm$1.4)$\times10^{-3}/K^-_{stop}$. 
Indeed, the quoted $\Gamma_{\Lambda d}$ is the intrinsic width of the 
bump which, when summed in quadrature with the bin width ($\sigma_{bin}$
=5.8 MeV/c$^2$) and spectrometer resolution, ($\sigma_{\Lambda d}$=6.0 
MeV/c$^2$), yields the full width of the Gaussian distribution.  

When the experimental data are fitted to a linear combination of 
only (1), (2) and (3) phase spaces (curve not shown), the chi-square 
($\chi^2_{123}$) is 7.61 with 5 degrees of freedom. The inclusion of $G$ 
yields $\chi^2_{123G}$=0.17 with 2 degrees of freedom. Since 
$\chi^2_{123}/5\gg\chi^2_{123G}/2$, the additional term $G$ significantly 
improves the goodness of the fitting. A measure of this improvement is 
given by the $F$ test for validity of adding the $G$ term,  where 
$F=\frac{2}{3}(\chi^2_{123}-\chi^2_{123G})/(\chi^2_{123G})$. In this case 
$F$=29.2. This large value of $F$ ensures a confidence level  slightly 
above 95\% in the relative merit of the $G$ term \cite{math:bevington}.
 
The statistical significance $\mathcal{Z}$ of the $m_{\Lambda d}$=3250 
MeV/c$^2$ signal is evaluated by using the  Uniformly Most Powerful 
test among the class of Unbiased tests (UMPU) method, which is 
described in Ref.\cite{math:statsig}. In this scheme, the total number 
of events observed around the bump is evaluated at 3$\sigma$. In the 
same 3$\sigma$ interval, the number of background events is accounted 
for by the number of events below the full-line curve. The 
resulting statistical significance is $\mathcal{Z}_{3\sigma}$= 3.9.  

The cos$\Theta_{\Lambda d}$ distribution of these bump events falling 
in the mass interval 3220$\le m_{\Lambda d}\le$3280 MeV/c$^2$ is shown
in the inset of Fig. 3. The distribution peaks at cos$\Theta_{\Lambda d}$
=-1 and spans a few bins above it. Its narrow width indicates that the 
$\Lambda d$ events arise from slowly-moving clusters. This is corroborated 
by the momentum distribution of $\Lambda d$ pairs ($p_{\Lambda d}$) in the 
mass range of the bump (figure not shown), the average value of which is
about 190 MeV/c (4.7 MeV), which corresponds to a particle with 
$m_{\Lambda d}$=3251 MeV/c$^2$ moving with a $\beta$=0.058.  Below the
 bump region $m_{\Lambda d}\le$3220 MeV/c$^2$, the cos$\Theta_{\Lambda d}$ 
distribution (figure not shown) is still peaked at -1 but with a FWHM about 
4 times broader than the cos$\Theta_{\Lambda d}$ distribution of the bump 
events.  

In the study of $^6Li$, the $\Lambda d$ channel leaves an undetected $3N$ 
system which can be a $t$, $nd$ or $nnp$ in the reaction final state. For 
$nd$, the  missing kinetic energy  distribution ($dn/dT_{mis}$)  can be 
determined  by means of the equation 
$T_{mis}=(m_{K^-_{stop}}+ m_{^6Li}-m_\Lambda - m_n -2m_d )-(T_\Lambda+T_d)$,
where ($T_{\Lambda}$ + $T_d$) is the sum of the kinetic energies of the 
detected particles, and $m$ is the mass of a generic  particle. The 
measured  distribution of the missing kinetic energy is shown in Fig. 4 
(thin-line histogram). It is shown only in the region up to 100 MeV where 
intensities are more accurately known. There is a bump in the kinetic energy  
%
%
\begin{figure}[t,c,b]
 \centering
  \includegraphics*[angle=90,width=0.5\textwidth]
   {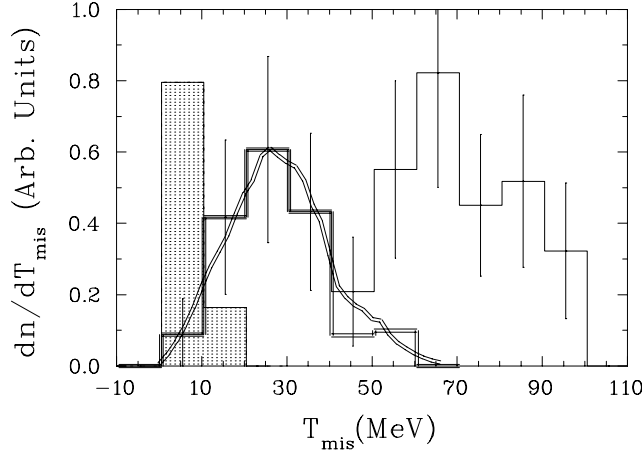}
    \caption{\footnotesize   Missing kinetic energy distribution of 
      the $^6Li(K^-_{stop},\Lambda d)nd$  reaction (thin-line histogram). 
      Thick-line histogram, $T_{mis}$ distribution for events correlated 
      to the $m_{\Lambda d}$=3250$\pm$30 MeV/c$^2$ mass range. Thick-line 
      curve,  simulated missing kinetic energy distribution of the 
      undetected $nd$ pairs with the energy of the spectator deuteron 
      below 4 MeV. The curve is normalized to the thick-line experimental 
      data. Grey-fill histogram,  simulated $T_{mis}$ distribution for 
      the $^6Li(K^-_{stop},\Lambda d)t$ reaction correlated to the 
      $m_{\Lambda d}$=3250$\pm$30 MeV/c$^2$ mass interval. The histogram is 
      arbitrarily normalized to the data. More details are given in the text.}
\end{figure}      
distribution at around 25 MeV which is strongly correlated to the 3250 
MeV/c$^2$ bump of Fig. 3; in fact, a 3220$\le m_{\Lambda d}\le$3280 MeV/c$^2$ 
constraint produces the thick-line histogram of Fig. 4. Such an occurrence 
can be explained by assuming the absorption process to start with the 
$\alpha (K^-_{stop},\Lambda d)n$ reaction, where $\alpha$ is a substructure 
of $^6Li(\equiv \alpha + d_H)$ \cite{expt:FINUDA2}. The $d_H$ deuteron of 
$^6Li$ is assumed to participate in the transition only by obeying the 
Hulthen momentum distribution \cite{theor:hulthen} which lies between 0 
and 150 MeV/c being peaked at about 50 MeV/c. To account for the kinematics 
of the full reaction, the simulated distributions of $\Lambda d$ events are 
shaped to fit the measured distributions in the ranges 
$p_{\Lambda ,d}$=600$\pm$150 MeV/c, 3220$\le m_{\Lambda d}\le$3280 MeV/c$^2$ 
and -1$\le$cos$\Theta_{\Lambda d}\le$-0.9. The result of these simulations 
is the thick-line curve in Fig. 4, which is normalized to the thin-line
experimental data. The curve  follows closely the behavior of the data, 
which supports the assumption that the initial stage of the reaction 
is the $K^-_{stop}\alpha$-cluster absorption, and most of the kinetic 
energy is taken away by the undetected neutrons. 

 For the $^6Li(K^-_{stop},\Lambda d)t$ reaction, the simulated 
$dn/dT_{mis}$ distribution is represented by the grey-fill histogram, 
which is arbitrarily normalized to the data. The constraint 
3220$\le m_{\Lambda d}\le$3280 MeV/c$^2$ requires tritons to peak at 
about 10 MeV kinetic energy. Such a peak does not appear in the 
experimental missing kinetic energy distribution, denoting that 
only limited strength is available to the $\Lambda d t$ channel. As well, 
the $\Lambda d ppn$ channel plays little part in building the 25 MeV bump 
because of its negligible strength in the $m_{\Lambda d}$=3250$\pm$30 
MeV/c$^2$ mass interval (i.e. dashed-line curve of Fig. 3). As a final 
comment, $\Sigma^0\rightarrow\Lambda\gamma$ decays can only feed 
the high-energy region of the $T_{mis}$ spectrum since the  undetected 
$\gamma$-rays take away 74 MeV.

The method employed  to analyze the $K^-$ absorption in $^6Li$ was 
also employed  to analyze the $^{12}C(K^-_{stop},\Lambda d)A'$ data. 
In general, these data are spread over a larger angular interval; for 
instance, compare the distribution of Fig. 3 inset with that of Fig. 
5 inset. This behavior is probably due to $\Lambda-A'$ and $d-A'$ 
final state interactions (FSI); in fact, FSI for $^6Li$ are smaller 
than for $^{12}C$ since the  residual nucleus is simply a deuteron. 
Fig. 5 shows the distribution of the $\Lambda d$ invariant mass of 
$^{12}C$. The high-energy part resembles a shoulder in shape, which
%
%
\begin{figure}[b]
 \centering
  \includegraphics*[angle=90,width=0.6\textwidth]
   {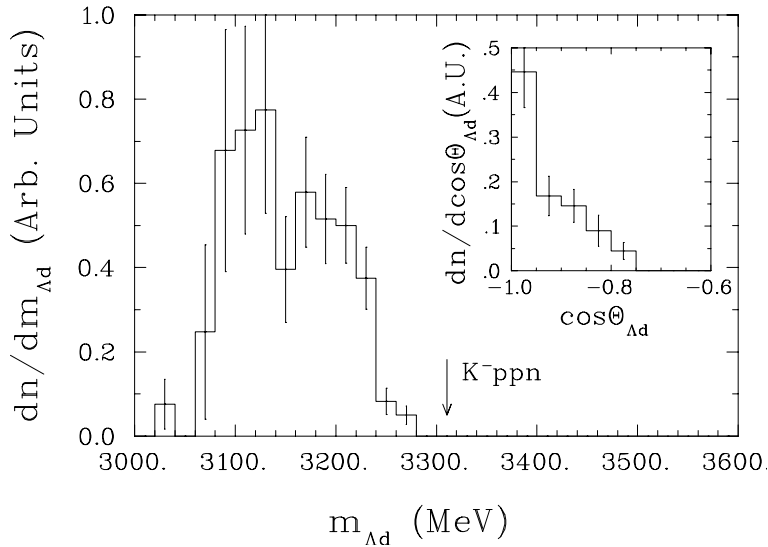}
    \caption{\footnotesize  Invariant mass distribution of $\Lambda d$ 
      pairs  from the $^{12}C(K^-_{stop}, \Lambda d)A'$ reaction. 
      Inset, cos$\Theta_{\Lambda d}$ distribution when  the invariant 
      mass of the $\Lambda d$ pairs is cut between 3220 and 3280 
      MeV/c$^2$.}
\end{figure}      
cannot be uniquely  fit with a Gaussian distribution as was done 
for $^6Li$. The cos$\Theta_{\Lambda d}$ distribution for the 
3220$\le m_{\Lambda d}\le$3280 MeV/c$^2$ events is reported in the inset 
of Fig. 5. This distribution, as well as the similar distribution for 
$^6Li$ (inset of Fig. 3), is peaked at around -1 although now it has a 
larger width. The $m_{\Lambda d}$ behavior was also examined for 
cos$\Theta_{\Lambda d}\le -$0.9, the  same angular range as $^6Li$. 
However, the data do not display any evident  bump-like structure.
\begin{center}
{\bf IV. DISCUSSION}   
\end{center}
The results were obtained by comparing the behavior of three
observables; i.e., $m_{\Lambda d}$, cos$\Theta_{\Lambda d}$ and
$T_{\Lambda d}$ the last two being independently measured. The $^6Li$ 
measurements provided results of primary interest in the  
$m_{\Lambda d}$=3250$\pm$30 MeV/c$^2$ mass range whose interpretation 
required extensive modeling. The $^{12}C$ data appear to be affected 
by FSI, which partially screens the underlying physics.  

For kaon absorption in $^6Li$, reaction kinematics were required 
to feed both the 3250$\pm$30 MeV/c$^2$ mass range and the correlated 
25$\pm$25 MeV kinetic energy range. The $\Lambda dnd$ reaction 
channel meets these requirements along with a moderate $\Lambda dnnp$ 
contribution. The monotonic behavior of the $\Lambda dnd$ and 
$\Lambda dnnp$ phase  spaces in the mass interval 3250$\pm$30 MeV/c$^2$ 
cannot explain the bump structure. Both the  $\Lambda dt$ channel and 
the $\Sigma^0\rightarrow\Lambda\gamma$ decay bring a negligible 
strength in the $m_{\Lambda d}$=3250$\pm$30 MeV/c$^2$ interval, whereas 
multistep processes are disfavoured by the reaction kinematics. 
Therefore, the symmetric shape of the bump was fit with a Gaussian 
distribution, which yields $m_{\Lambda d}$=3251$\pm$6 MeV/c$^2$, 
$\Gamma_{\Lambda d}$=36.6$\pm$14.1 MeV/c$^2$ and $\mathcal{Z}_{3\sigma}$
=3.9.  Since the bump at about 3140 MeV/c$^2$ is probably an artifact 
due the poor statistics along with the low acceptance of the apparatus, 
it cannot be compared with a similar structure at $m_{\Lambda d}$=3160 
MeV/c$^2$ obtained by the FOPI collaboration in heavy ion measurements 
\cite{expt:FOPI}.

Extensive $K^-_{stop}$$^6Li\rightarrow \Lambda d nd$ modeling  which 
relied on the cluster structure of lithium, $^6Li(\equiv\alpha + d_H)$ 
\cite{expt:FINUDA2} was required to describe the $T_{mis}$=25 MeV bump 
(Fig. 4). In fact, a reaction mechanism capable of explaining the bump 
requires that negative kaons are preferentially absorbed by $\alpha$-like 
substructures, $K^-_{stop}\alpha\rightarrow [K^-ppn] + n$, while the $d_H$ 
deuterons participate in these processes as spectators. Final state 
neutrons remove the excess energy. In these dynamics, neutrons are the 
{\em spectroscopy particles} of the $[K^-ppn]$ cluster formation. The 
$[K^-ppn]$ clusters finally decay via the $[K^-ppn]\rightarrow\Lambda d$  
channel. $\Lambda$-hyperons and deuterons are observed to have a strong 
angular correlation at around 180$^\circ$. In this measurement, other 
decay channels were not examined. 

$[K^-3N]$ clusters were discussed earlier  in the framework of 
$\overline{K}$ nuclear bound states. The nuclear ground state of 
[$K^-\otimes$ $^3He$ + $\overline{K}^0\otimes$ $^3H$] was predicted to 
be 108 MeV deep and 20 MeV wide \cite{theor:akaishi}. Similar quantities 
can be determined by assuming [$\overline{K}$$\,ppn$]$\equiv$
[$K^-\otimes$ $^3He$ + $\overline{K}^0\otimes$ $^3H$], such as the 
$[K^-ppn]$ binding energy $B_{K^-ppn}$=$(m_{K^-}+2m_p+m_n)-m_{\Lambda d}$
=58$\pm$6 MeV and $\Gamma_{\Lambda d}$=36.6$\pm$14.1 MeV/c$^2$, where $m$ 
is the generic particle rest mass.  Although the theoretical predictions 
agree with experiments supporting formation of kaonic nuclear states, the 
agreement  is poor.  A comment on the binding energy and width is 
worthy of note. The values of $B$ and $\Gamma$ depend on the bump position 
and width, respectively. In this measurement, the bump develops close to 
the phase space endpoint of the $^6Li(K^-_{stop},\Lambda d)nd$  reaction
(3280 MeV, see Fig. 3), which  might modify the structure of the bump thus 
 $B_{K^-ppn}$  and $\Gamma_{\Lambda d}$.

A recent study of the $K^-_{stop}A\rightarrow\Lambda pA'$ reaction, where 
$A$ combines 3 targets $^6Li,^7Li$ and $^{12}C$ \cite{expt:FINUDA0}, found 
evidence of a kaon bound state $[K^-pp]$. The binding energy $B_{K^-pp}$=
115 MeV and the decay width $\Gamma_{\Lambda p}$=67 MeV/c$^2$ were determined 
from the $\Lambda p$ invariant mass distribution. For the case of the 
$\Lambda p$ measurement, both parameters $B$ and $\Gamma$ are larger than 
our results. The result seems to contradict a theoretical expectation for 
small binding energy to be correlated with large decays widths. In view of 
these results, further theoretical and experimental developments are needed. 
The  existence of bound kaonic states in nuclei remains controversial 
\cite{theor:oset1,theor:baca,theor:mares}, and needs to be definitively 
settled.  

The present analysis indicates an alternative way to detect   
[$\overline{K}ppn$] cluster decays when forming in $^4He$ or $ ^6Li$. 
Formation of these clusters is accompanied by  neutrons whose kinetic
energy is below 50 MeV. If the role of FSI is marginal, these neutrons 
should peak at around 25 MeV (Fig. 4 thick-line). For $^{12}C$, 
such a distinctive pattern disappears (spectrum not shown) probably 
being engulfed by the continuum.  Proposals to use the $(K,n)$ reaction 
to search  for strange heavy baryons\cite{theor:kishimoto} should account 
for these results. 
\begin{center}
{\bf V. CONCLUSIONS}   
\end{center}
We report in this letter the results of a kaon absorption study  by means 
of the $K^-_{stop} A\rightarrow\Lambda d \, A'$ reaction; the nuclei 
examined were $^6Li$ and $^{12}C$. The experiment was performed with the 
FINUDA spectrometer, which was installed at the DA$\Phi$NE $\phi$-facility 
(LNF). The study makes full use of the capability of FINUDA to reconstruct 
the tracks of most particles involved in the absorption reaction with the 
exception of the residual nucleus. The analyses deal with correlated 
particle pairs $\Lambda d$, which considerably clean the spectra from 
accidental events.

For light nuclei, the $K^-$ absorption reaction  presents some common 
features:  (1) The m$_{\Lambda d}$ distribution develops its strength 
well below the sum of the $K^-, p, p, n$ rest masses (denoted with an 
arrow in Figs. 3 and 5). (2) A strong angular correlation exists between 
$\Lambda$ hyperons and deuterons; in fact, the $\Theta_{\Lambda d}$ 
distribution results are sharply peaked at  around 180$^\circ$.       

In $^6Li$, a bump was found at the tail of the $m_{\Lambda d}$ 
distribution with a peak at 3251$\pm$6 MeV/c$^2$, a width 36.6$\pm$14.1 
MeV/c$^2$, and a yield of (4.4$\pm$1.4)$\times10^{-3}/K^-_{stop}$ with a  
statistical significance 3.9.  The combined picture of the opening angle 
distribution of the detected $\Lambda d$ pairs and their momentum 
distribution shows that the $\Lambda d$ events are emitted from [$K^-ppn$] 
slow-moving clusters. The cluster momentum is counterbalanced by $T_n <$
50 MeV neutrons.  The corresponding $[K^-ppn]$ binding energy is 
$B_{K^-ppn}$=58$\pm$6 MeV. In the case of $^{12}C$, a shoulder structure 
replaced the bump in the $^6Li$ $\Lambda d$ invariant mass spectrum. An 
analysis based on  statistical methods does not yield a unique set of  
$m_{\Lambda d}$ ($B_{K^- ppn}$) and $\Gamma_{\Lambda d}$ parameters.   

The values of $B_{K^- ppn}\sim$58 MeV and $\Gamma_{K^-ppn}\sim$37 
MeV/c$^2$ obtained for the [$\overline{K}ppn$] bound system can be 
compared to the values of an earlier theoretical work
\cite{theor:akaishi} based on the possible existence of $\overline{K}$ 
bound nuclear states. Based on a ground state of  
[$K^-\otimes$ $^3He$ + $\overline{K}^0\otimes$ $^3H$], $B_{K^-}$=108 MeV 
and $\Gamma_{K^-}$=20 MeV. However overall agreement is poor and improved 
calculations are needed, especially now that  the present result is 
consistent with the formation of a bound [$K^-ppn$] cluster, and the 
reaction mechanism for neutrons has been  deduced. Increasing 
theoretical interest in obtaining a reliable physical framework for 
analysis of recent data is evidenced by the number of recent publications 
\cite{theor:mares,theor:zhong,theor:dote,theor:ivanov,theor:weise}. 

%
%
%
 
%
%
\end{document}